\documentclass[10pt,letterpaper,conference]{IEEEtran}
\IEEEoverridecommandlockouts

\usepackage{cite}
\usepackage{amsmath}
\usepackage{amssymb}
\usepackage{adjustbox}
\usepackage{algorithmic}
\usepackage{fixltx2e}
\usepackage{graphicx}
\graphicspath{ {images/} }

\usepackage{kbordermatrix}

\usepackage{enumerate}
\usepackage{soul}  
\soulregister\cite7 

\usepackage{url}
\usepackage{algorithmic}
\usepackage{subfigure}
\usepackage{algorithm}
\usepackage{multirow}
\newlength\myindent
\usepackage{tikz}
\usetikzlibrary{shapes, arrows, decorations, decorations.markings, backgrounds, calc, arrows.meta, shadows, shadows.blur,positioning,spy,shadings}
\pgfdeclarelayer{myback}
\pgfdeclarelayer{myback2}
\pgfsetlayers{background,myback2,myback,main} 

\usepackage{pgfplots}
\pgfplotsset{width=10cm,compat=newest}
\pgfplotsset{compat=newest} 
\pgfplotsset{plot coordinates/math parser=false} 
\newlength\figureheight 
\newlength\figurewidth 
\usepgfplotslibrary{units}
\usepackage{pgfplotstable}

\tikzstyle{vecArrow} = [{line width=0.5pt, double distance=5pt, arrows={-Implies[length=0pt 0.8 0]}}]
\tikzstyle{vecNoArrow} = [anchor=center,decoration={markings,mark=at position
	1 with },
double distance=5pt, shorten >= 3pt,
preaction = {decorate},
postaction = {draw,line width=1.5pt, white,shorten >= 1.5pt}]
\tikzstyle{vecArrowTUMblue} = [anchor=center,decoration={markings,mark=at position
	1 with {\arrow{Straight Barb[length=6pt,TUMblue]}}},
double distance=5pt, shorten >= 3pt,
preaction = {decorate},
postaction = {draw,line width=1.5pt, white,shorten >= 1.5pt}]
\tikzstyle{innerWhite} = [semithick, white,line width=4pt, shorten >= 4.5pt]
\tikzset{
	triangle/.style = { regular polygon, regular polygon sides=3, rotate=-90,thick,scale=1.2},
	triangleWithText/.style = { regular polygon, regular polygon sides=3, shape border rotate = -90, rotate=0,thick,scale=1.2},
	sum/.style = {draw, circle,thick,scale=1.3},
}

\definecolor{mycolor1}{rgb}{0.00000,0.44700,0.74100}%
\definecolor{mycolor2}{rgb}{0.85000,0.32500,0.09800}%
\definecolor{mycolor3}{rgb}{0.92900,0.69400,0.12500}%
\definecolor{mycolor4}{rgb}{0.49400,0.18400,0.55600}

\usepgflibrary{patterns} 
\usepgflibrary[patterns] 
\usetikzlibrary{patterns} 
\usetikzlibrary[patterns] 

\usepackage{textcomp}
\newcommand\copyrighttext{%
	\footnotesize \textcopyright 2019 IEEE.  Personal use of this material is permitted.  Permission from IEEE must be obtained for all other uses, in any current or future media, including reprinting/republishing this material for advertising or promotional purposes, creating new collective works, for resale or redistribution to servers or lists, or reuse of any copyrighted component of this work in other works.
}
\newcommand\copyrightnotice{%
	\begin{tikzpicture}[remember picture,overlay]
	\node[anchor=south,yshift=10pt] at (current page.south) {\fbox{\parbox{\dimexpr\textwidth-\fboxsep-\fboxrule\relax}{\copyrighttext}}};
	\end{tikzpicture}%
}

\newcommand{\journal}[1]{#1} 
\renewcommand{\journal}[1]{} 

\large{\title{VRLS: A Unified Reinforcement Learning Scheduler for Vehicle-to-Vehicle Communications}}

\author{
	\IEEEauthorblockN{
			Taylan \c{S}ahin\IEEEauthorrefmark{1}\IEEEauthorrefmark{2}, 
			Ramin Khalili\IEEEauthorrefmark{1},
			Mate Boban\IEEEauthorrefmark{1} and
			Adam Wolisz\IEEEauthorrefmark{2}
		}

		\IEEEauthorblockA{
			\IEEEauthorrefmark{1}Huawei Technologies Duesseldorf GmbH, 80992 Munich, Germany\\
			Email: \{taylan.sahin, ramin.khalili, mate.boban\}@huawei.com
		}
		
		\IEEEauthorblockA{
			\IEEEauthorrefmark{2}Telecommunication Networks Group, Technische Universit{\"a}t Berlin, 10587 Berlin, Germany\\
			Email: adam.wolisz@tu-berlin.de
		}
}

\begin{document}
	\maketitle
	\copyrightnotice  
	
	\begin{abstract}

Vehicle-to-vehicle (V2V) communications have distinct challenges that need to be taken into account when scheduling the radio resources. 
Although centralized schedulers (e.g., located on base stations) could be utilized to deliver high scheduling performance,
they cannot be employed in case of 
coverage gaps. 
To address the issue of reliable scheduling of V2V transmissions out of coverage, we propose Vehicular Reinforcement Learning Scheduler (VRLS), 
a centralized scheduler that predictively assigns the resources for V2V communication while the vehicle is still in cellular network coverage.
%


VRLS is a unified reinforcement learning (RL) solution, wherein the learning agent, the state representation, and the reward provided to the agent are applicable to different vehicular environments of interest (in terms of vehicular density, resource configuration, and wireless channel conditions). Such a unified solution eliminates the necessity of redesigning the RL components for a different environment, and facilitates transfer learning from one to another similar environment.


We evaluate the performance of VRLS and show its ability to avoid collisions and half-duplex errors, and to reuse the resources better than the state of the art scheduling algorithms. We also show that pre-trained VRLS agent can adapt to different V2V environments with limited 
retraining, thus enabling real-world deployment in different scenarios. 

\end{abstract}

	
	\section{Introduction}
\label{Introduction}


Achieving high reliability of vehicle-to-vehicle (V2V) communication is a challenging task, since the highly dynamic vehicular environment is prone to collisions due to, e.g., hidden node problem and simultaneous transmissions.
Combined with the half-duplex constraint~\cite{ding2012combating} in case of {\color{black} resource scheduling in both time and frequency},
the task of V2V resource allocation becomes particularly challenging. Recent efforts have focused on enabling V2V communication using cellular networks, under the name of C-V2X \cite{3gppTR36885}, \cite{3gppTR38885}. 
While C-V2X can operate without infrastructure, one of the main benefits it brings is a highly efficient coordination of V2V transmissions via the network (i.e., through base stations). However, highly reliable V2V transmissions should not rely on ubiquitous availability of network coordination, since there will exist coverage gaps, either because of physical impediments (e.g., tunnels, blockage of link by large objects such as buildings, etc.) or due to infrastructure deployment, which might be insufficient to cover the entire roadway. To support advanced V2V use cases \cite{3gppTR22886}, highly reliable V2V communication needs to be enabled both in-coverage and out-of-coverage (OOC).

In our initial work on the topic \cite{sahin2018radio},\cite{sahin2018reinforcement}, we scoped the problem of scheduling V2V communications in OOC, and explored the ability of the centralized reinforcement learning (RL) scheduler to ``pre-schedule'' the V2V transmissions for OOC. We have shown that there lies a strong promise in using RL to efficiently schedule periodic V2V transmissions for OOC areas on highway that experience different vehicular and data traffic.

On the other hand, RL has its own domain-specific challenges that require careful consideration, especially 
in pursuance of practical solutions. For a specific problem, design choices of the RL components (the learning agent, the state information, and the reward signal) play a paramount role, as they heavily affect the performance \cite{sutton1998reinforcement}. However, although RL solutions are traditionally designed for, trained, and evaluated on the same environment, such approach turns into a limitation in practice. Recent work shows RL agents tailored for a specific environment may not perform as well on, or be applicable at all to, even a slightly different environment \cite{zhao2019investigating}.


In this paper, we propose Vehicular Reinforcement Learning Scheduler (VRLS), a unified RL scheduling approach that is applicable  to a variety of vehicular environments of interest. VRLS is designed as a centralized scheduler (e.g., one residing in the network) consisting of an RL agent taking actions (assignment of resources to the vehicles), and the state representation and reward definition provided to it. The structure of all three components of VRLS remain the same, irrespective of what kind of setting they are applied to, which allows for a broad applicability of VRLS to different practical scenarios.
Our main contributions are as follows:
\begin{itemize}
\item We design VRLS, an RL scheduler applicable to different configuration of resources in time and frequency, vehicle densities, and channel conditions, thanks to its unified approach (Section \ref{Algorithm}).
\item {\color{black} VRLS further improves the performance of scheduling the V2V communications, in terms of reduced packet error rates achievable in OOC, compared to state of the art algorithms~\cite{3gppTR36885},~\cite{sahin2018reinforcement} (Section \ref{Results}).}
\item With limited retraining, the learning performed by VRLS over simplified environments can be transferred to more realistic, complex environments.
\end{itemize}

\begin{figure}[!t]
	\centering
	\includegraphics[width=\columnwidth]{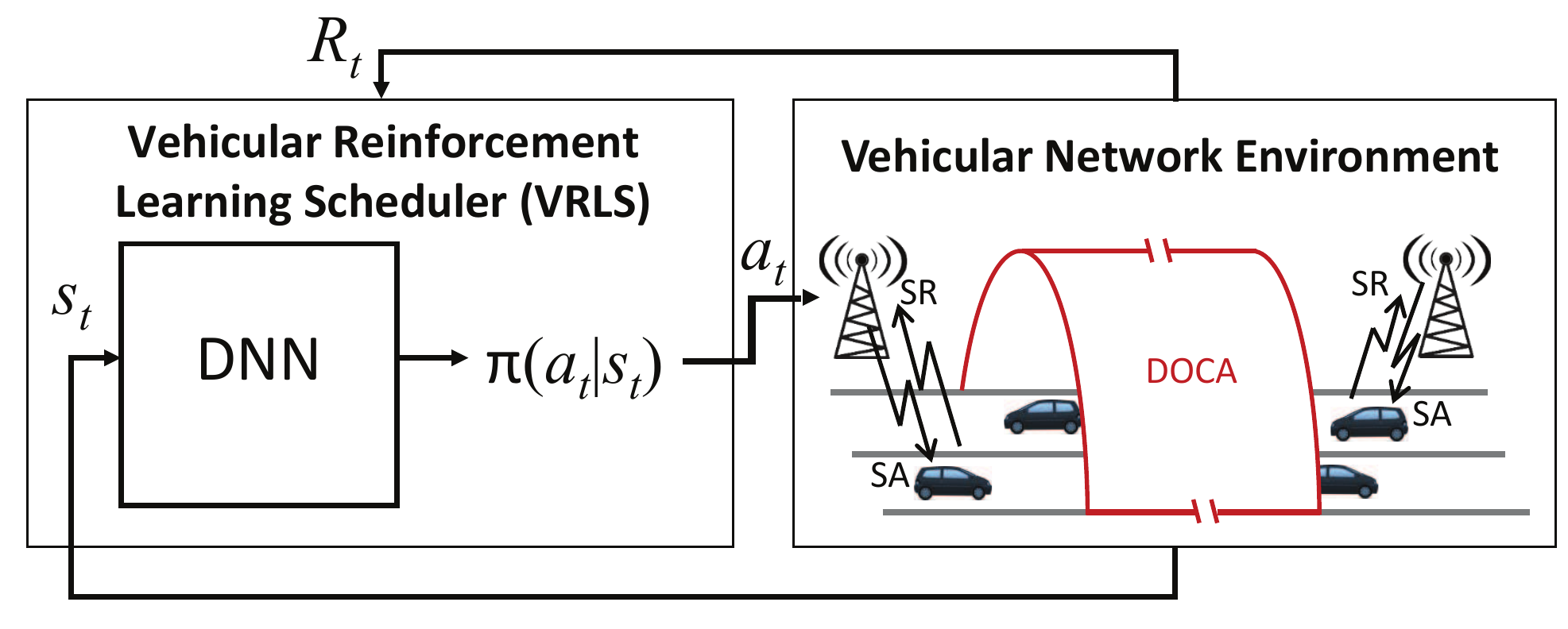}
	\caption{RL applied to our vehicular network environment, DOCA (delimited out-of-coverage area). Vehicles communicate with each other using the scheduling assigments (SA) sent by the delimiting BSs upon their scheduling request (SR) before they enter the area. VRLS is responsible for the reliability of V2V messages in DOCA.}
	\label{RL}
\end{figure}

The rest of the paper is structured as follows. Section~\ref{Model} provides the system model and defines the problem. Section~\ref{Algorithm} describes VRLS, Section \ref{Results} presents the evaluation results, and Section \ref{Conclusion} concludes the paper.
	
	\section{System Model}
\label{Model}

\subsection{Vehicular Network Environment}
We consider V2V communications in out-of-coverage (OOC), where radio resources need to be assigned so that high reliability can be achieved. Unlike the existing approaches (e.g., 3GPP Mode 4~\cite{3gppTR36885}), we rely on a centralized scheduler to ``pre-schedule'' the resources for OOC while vehicles are in coverage. 
While in network coverage, vehicles 
receive the scheduling assignments (SAs) from a centralized scheduler for OOC transmissions through a base station (BS). As we are interested in coverage gaps, we consider a delimited out-of-coverage area (DOCA) of the network, which is illustrated in Fig.~\ref{RL}. DOCA is delimited by the \textit{coverage} of the BSs surrounding it, e.g., a segment of a highway remaining outside the coverage of BSs at its two ends, as shown in Fig.~\ref{RL}.  
That is, BSs are still able to provide SAs to the vehicles for their transmissions that will take place in DOCA, before vehicles enter it, i.e., go out of their coverage, irrespective of which direction in DOCA the vehicles take. Accordingly, BSs are assumed to be aware of the existence and topology of DOCA (e.g., via the network operator or by measuring the signal level and detecting the link interruptions). The V2V transmissions are assumed to consist of periodic broadcast messages (e.g., cooperative awareness messages: CAMs~\cite{ETSIEN3026372}), each occupying a certain amount of time-frequency resources. 


\subsection{Time-Frequency Resources}

\begin{figure}[!t]
	\centering
	\includegraphics[scale=0.75]{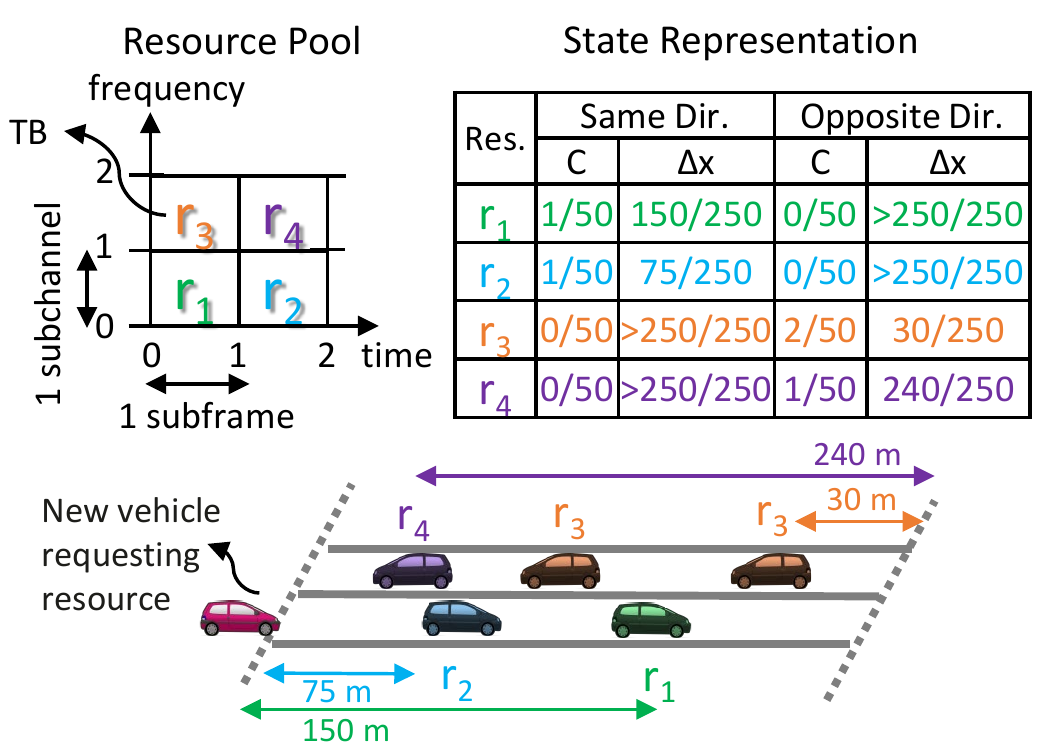}
	\caption{Radio resource pool (top left) and RL state representation (top right) for an exemplary scenario (bottom).}
	\label{Pool}
\end{figure}

V2V communication utilizes a dedicated pool of time-frequency resources which is configured by the network with a certain number of subchannels (over frequency dimension) and subframes (over time dimension). Fig.~\ref{Pool} illustrates an example resource pool configured with $2$ subframes and $2$ subchannels. Following the LTE-V assumptions~\cite{3gppTR36885}, a transmission of a CAM occupies a single transmission block (TB) that consists of a single subframe of $1$~ms and a single subchannel (containing sufficient number of resource blocks given the CAM size, and the modulation and coding scheme). Accordingly, the SA provided by the BS to the vehicles entering DOCA indicates which TB to use in the next available resource pool, each time they generate CAMs during their travels in DOCA. In this way, the maximum latency of CAMs is also limited by the time-length of the pool, i.e., the number of subframes. The resource pool is assumed to be configured exclusively for DOCA, i.e., not shared with in-coverage vehicles, hence precluding any interference from them. However, this is simply a design choice for the system model, and VRLS can be used without modifications in case of resources shared between vehicles that are in DOCA and those in coverage.

Vehicles transmitting at the same time, i.e., using TBs in the same subframe, are not able to receive each other's message. This is the well known half-duplex (HD) limitation for radios that can either transmit or receive (but not both) at a certain time~\cite{ding2012combating}. We refer to this situation at the receiver side as \textit{HD error} or \textit{conflict}, and refer to the relation among the TBs in the same subframe leading to this phenomenon, as \textbf{\textit{HD constraint}}, in the rest of the paper. Moreover, messages transmitted from different vehicles in the same subframe \textit{and} the same subchannel, i.e., using the same TB, may, depending on propagation conditions, interfere or even collide completely with each other, leading to decoding errors at the receivers. Accordingly, we refer to such situation resulting in unsuccessful reception as \textbf{\textit{collision errors}}.

Regarding the propagation on the wireless channel, we consider two conditions on the vehicular environment: i) a DOCA of single collision domain (SCD), where all of the vehicles are within transmission range of each other, and cause significant interference to each other if transmitting at the same time; and ii) a DOCA of multiple collision domains (MCD), where the effective transmission range of the vehicles are smaller than the DOCA size. The former results in collision errors at all receivers in case the same TB is used by more than one transmitting vehicle. Similarly, HD errors would occur among all the vehicles using the TBs in the same subframe. {\color{black} We use SCD environment as a ``sanity-check'' in our evaluations, since it enables an analytically tractable estimation of optimum performance of a scheduler.}
On the other hand, MCD conditions the successful decoding of the message at the receiver on the signal-to-interference-plus-noise ratio, which depends on the transmitter-to-receiver distance, as well as the interference level from other transmissions using the same TB, when the pathloss and fading effects are taken into account. Accordingly, reuse of the same TB or use of the TBs sharing the same subframe is possible among transmitters sufficiently far apart, without creating any collision and HD errors at the receivers, respectively.

\subsection{Problem Definition}

Our objective is to ensure reliability of the V2V transmissions in DOCA by pre-scheduling TBs to vehicles, as they go out of coverage. Reliability is quantified using the metric from the 3GPP standard~\cite{3gppTR36885}, called average packet reception ratio (PRR). PRR is defined as the ratio of the number of vehicles having successfully received one transmitted message to the total number of vehicles in the range of interest. The average is then calculated for all transmissions performed within a certain time interval by dividing the sum of the number of vehicles successfully receiving each transmission to the sum of the number of all vehicles within the specific range of each transmission. 

Achieving reliability in DOCA is a challenging task, considering the limited amount of resources to be managed for highly dynamic nodes. The resources should be allocated in such a way as to not create interference or collisions at the receivers, as well as to avoid HD conflicts. 

\subsection{RL-based Approach and its Challenges}
We deal with the objective of providing reliable V2V communications in DOCA by using an RL-based scheduling approach. Our solution
involves an \textit{RL agent}, a logically centralized entity residing in the network (e.g., at the BSs), which schedules the V2V transmissions in DOCA. The agent takes an \textit{action} $A_t$, at each time $t$ a vehicle enters DOCA. The action we define is to assign the entering vehicle a single TB for its V2V transmissions through DOCA. The decision on which TB to be assigned at time $t$ is given by the agent's \textit{policy} $\pi : \pi(a_t | s_t) \rightarrow[0,1]$, which is a mapping from the agent's current observations from its environment, i.e., \textit{state} $S_t$ of DOCA (defined in Section \ref{State}), to a probability distribution over the set of possible actions, i.e., the TBs in the resource pool.
We model the policy using a \textit{deep neural network} (DNN), whose parameters are trained by the agent solely through its interaction with the environment. Training is conducted by providing the agent upon its each action a \textit{reward} signal $R_{t+1}$ from the environment (defined in Section \ref{Reward}), which quantifies the fitness of the selected action, hence the policy. The evolution of this signal determines how the agent should update its policy. Updates are performed according to the specific RL algorithm (described in Section \ref{TrainingAlgorithm}), where the agent's goal is to maximize the cumulative reward it receives on the long run. 


In our previous work \cite{sahin2018reinforcement}, we observed that, each time we consider a different vehicular environment, we were required to revise our RL design to reach a desirable performance. In other words, we modified the state representation, reward definition, as well as the structure of the underlying DNN, when the environment changed considerably (e.g., in terms of vehicle density, speeds, or the structure of the resource pool), to guarantee convergence to a ``good'' policy through learning. Such an approach is impractical for arbitrarily different new environments, 
and 
the policy learned on one environment cannot be used as a starting point in another environment. Existing literature indicates that this limitation is not specific to our application of RL to V2V, but rather a general problem in deep learning across different domains. Specifically, recent studies report the challenge of deep RL, where standard algorithms and architectures are shown to perform poorly in case changes (e.g., noise) applied to the environment \cite{zhao2019investigating}. For example, authors of \cite{gamrian2018transfer} show that the famous deep Q-network mechanism \cite{mnihAtari}, which is trained to play Atari games and shown to outperform human level, fails completely when simple modifications are applied to the environment (e.g., adding pixels to the screen). Various approaches are proposed by these recent works, such as changing the state representation, applying a different DNN architecture, or training the agent from scratch, in order to achieve applicability and the desired performance of the RL agent on different environments, as fine-tuning is not always effective (cf. \cite{gamrian2018transfer}, \cite{zhao2019investigating}, and their references). Currently, there is no single, robust RL solution that can cope with a broad set of environments and parameters~\cite{zhao2019investigating}.

Motivated by the state-of-the-art, as well as our previous experience in \cite{sahin2018reinforcement}, we propose VRLS, a unified RL approach for the problem of scheduling V2V communications under different vehicular environments, which is able to handle variations in terms of available resources, network load, mobility, and wireless channel. We train VRLS in different scenarios, and show that it can learn near optimal policies in all these cases, and that the learning performed over simplified environments can be transferred to more realistic, complex environments, without requiring much retraining.

	\section{VRLS: Vehicular Reinforcement Learning Scheduler}
\label{Algorithm}

In this section, we present VRLS by describing the learning algorithm employed, state representation to the agent, definition of reward, as well as the design of the VRLS agent itself.


\subsection{Training Algorithm}
\label{TrainingAlgorithm}

The large space of possible combinations of different number of resources and vehicles makes tabular learning methods infeasible for this problem \cite{sutton1998reinforcement}. This leads us to apply approximate solution methods, where we utilize two DNNs, one used to represent the policy, containing a set of adjustable parameters $w$, i.e., $\pi_w(a_t | s_t)$, referred to as the \textit{actor} network, and the other used to represent the state values, referred to as the \textit{critic} network. The value of a state is defined as the expected rewards in a long run starting from that state and following the policy $\pi_w$ onwards, and is used as a \textit{critic} when training the policy parameters. Thanks to the policy gradient theorem \cite{sutton1998reinforcement}, an exact expression on how the performance is affected by the policy parameters can be derived for such actor-critic methods, providing strong convergence properties.

We utilize the A3C algorithm \cite{Mnih2016} to train the parameters of the actor-critic DNNs. Our solution enables training multiple agents in parallel, each interacting with a different instance of the environment. Each agent reports its experience (i.e., an \textit{epoch} of state-action-reward sequences) to a central coordinator, which in turn updates the parameters of the DNNs of all agents, used for their policy and state values.

\subsection{State Representation}
\label{State}
We design the state representation of the environment $S_t$ provided to the RL agent to carry the information on how the resources are utilized at each time $t$ a vehicle is entering the DOCA. As shown  in Fig.~\ref{Pool}, $S_t$ has a matrix structure, with number of rows equal to the number of  resources in the resource pool, and with four columns providing the following information about each resource. 
The first column of the state representation contains the count $C$ of how many times each resource has been assigned to the vehicles traveling in the direction \textit{same} as the vehicle entering DOCA, normalized to the maximum number of vehicles that can ever fit DOCA in that direction. The maximum number is derived by dividing the DOCA length with the vehicle length. The second column contains the estimated traveled distance $\Delta x$ of the vehicle to which the resource was last assigned, from the entrance point of DOCA in the direction \textit{same} as that of the entering-vehicle, normalized to the total DOCA size. $\Delta x$ is calculated by multiplying the vehicle speed at the time the vehicle going out of coverage by the amount of time passed. 
The last two pairs of columns provide the same information as the first two, but on the direction \textit{opposite} to the vehicle entering DOCA, respectively. An example with $4$ resources, and a DOCA of length $250$m, where a maximum of $50$ vehicles can fit per direction and lane, is illustrated in Fig.~\ref{Pool}. 


$C$ indicates 
whether the resources are free and, if not, how much they are loaded (or reused) in each direction. 
$\Delta x$ indicates how ``safe'' it will be to reuse the same resource in the same direction. In addition, the information on vehicle density per direction could be attained by accounting the total allocation counts in the first and the third columns. Furthermore, using the first pair of columns for the same direction as that of the vehicle requesting resource, and the opposite for the second pair enables us to encode also the direction information of the vehicle entering DOCA in the state representation. It is important 
for the agent to know the direction of the vehicle to which the resource is assigned in case of a DOCA with multiple collision domains, since the reliability would be affected differently depending on the resource utilization in each direction. 

Note that our state representation is applicable to any number of resources as well as different vehicular environments in terms of the size of DOCA and the number of vehicles within, owing to the use of normalized state variables.

\subsection{Reward Definition}
\label{Reward}
During the training, we impart the main goal -- maximizing the reliability of transmissions taking place in DOCA -- to the RL agent, by incorporating the reliability metric measured in the environment into the reward signal $R_{t+1}$ upon each action $A_t$. The objective of RL agent is to maximize the reward it collects on the long run. Specifically, we define the reward as a linear function of PRR, i.e., $R_{t+1}=-10\times(1-min(PRR))$. The minimum is taken over the ranges of interest where the PRR is measured, and in the measurements, the transmissions of the vehicles traveling through DOCA since the last action are taken into account. In case no transmissions take place between two actions, which could happen, e.g., when two vehicles enter DOCA almost at the same time, 
we provide the reward of the previous action to the agent.

\subsection{Design of the VRLS Agent}
We have utilized convolutional neural network for the VRLS agent, motivated by its practical success and advantages in processing different types of data \cite{nature}. However, as opposed to its traditional application on 2D images, we work with 2D data with rows corresponding to an ordered set of resources, and columns providing information about each. Accordingly, we feed each column of the state separately into different 1D convolutional layers. Output of each layer is then merged and fed into a single 2D convolutional layer together, enabling the agent first to process the relations across the resources (rows), and then across the different information provided for each resource (columns), respectively at the first two layers, where $tanh$ is used as the activation function. Since the actor network in our setting provides the action probabilities (i.e., the probability of selecting each of the TBs) given the state representation at its input, its final layer consists of a fully connected layer with number of units equal to the number of TBs, utilizing $softmax$ activation function that produces the probability distribution over the units, i.e., TBs. In case of the critic network, which estimates the value of a given state at the input, the final layer is a single unit fully connected to the previous convolution network, outputting the single estimated value using a linear activation function.


Note that convolutional layers have a property where their output depends on the \textit{order} of the input data they process. Although this is useful for their most common applications, such as processing images, we want the policy of the agent, i.e., the DNN, to be independent of the order of resources presented in the state, in our case. That is, the policy learned by the agent should not depend specifically on, or be limited to, raster-scan ordering, since it is an arbitrary choice, but rather depend on the information provided about each resource. To tackle this issue, we resort to data augmentation methods, where we \textit{randomly shuffle} the order of resources, i.e., the rows in the state representation, each time we feed it to the agent. The ordering of the probabilities of selecting each resource at the output layer of the actor-DNN is changed in the same way. This way, the agent becomes invariant to any kind of ordering during the training. On the other hand, given such a ``shuffled'' representation, the agent will never be able to infer how the actual pool of resources is configured. For example, a pool of $20$ resources could be configured in $5$ subframes by $4$ subchannels, or in $10$ subframes by $2$ subchannels, as well (see Fig. \ref{HDpools}), and there is no explicit information in the state representation related to it. The configuration, however, is important to the agent as different configurations would create different HD constraints among the resources of the pool. A policy without this information would easily result in performance degradation, depending on the vehicular environment. Therefore, we come up with a simple modification to shuffling, where we first group the resources sharing the same subframe, then randomly shuffle the order of these \textit{groups of resources}. To illustrate using Fig. \ref{Pool}, we first group the resources sharing the same subframe as ${[r_1,r_3]}$ and ${[r_2,r_4]}$, and then provide the state information corresponding to a random ordering of these groups. 
This way, the convolutional network would still be invariant to the order of resources in time, while being able to respect the partially-preserved order which could be exploited to infer the HD constraints among the resources specific to the pool configuration.

	\section{Evaluation}
\label{Results}

\begin{table}[!t]
	\renewcommand{\arraystretch}{1.1} 
	\caption{Simulation Parameters}
	\label{tableScenario}
	\resizebox{\columnwidth}{!}{%
		\centering
		\begin{tabular}{|l|l|l|l|l|}
			\hline
			& \textbf{MCD} & \textbf{SCD-i} & \textbf{SCD-ii} & \textbf{SCD-iii}\\
			\hline
			Maximum number of vehicles & $30$ & $10$ & $4$ & $5$\\
			\hline
			\multirow{2}{*}{Resource pool configuration} & $2$ subchannels & $2$ sch. & $10$ sch. & $4$ sch.\\
			& $10$ subframes & $10$ sf. & $2$ sf. & $5$ sf. \\								   
			\hline
			\multirow{2}{*}{DOCA size} & \multicolumn{4}{l|}{500 m of a straight highway,}\\
			& \multicolumn{4}{l|}{1 lane per direction, $4$~m lane width}\\
			\hline
			Vehicle speed & \multicolumn{4}{l|}{$50$ km/h}\\
			\hline
			Vehicle drop & \multicolumn{4}{l|}{Poisson distribution with mean of 2.5-s distance}\\
			\hline 
			Transmission power & $-5$ dBm & \multicolumn{3}{l|}{$23$ dBm (the maximum value)}\\
			\hline 
			CAM size and periodicity & \multicolumn{4}{l|}{$190$ B, $100$ ms}\\ 
			\hline
			Mode 4 ``probResourceKeep'' & \multicolumn{4}{l|}{$0$}\\
			\hline
			Number of actions per epoch & \multicolumn{4}{l|}{$60$}\\ 
			\hline
			Actor-critic learning rates & \multicolumn{4}{l|}{$10^{-3} / (1 + 0.01\times \#ep^{1.1})$}\\ 
			\hline
			\multicolumn{5}{|c|}{\textbf{V2V channel model in Fig. \ref{Results_E2} \cite{3gppTR36885}}}\\
			\hline
			\multirow{2}{*}{Pathloss model} & \multicolumn{4}{l|}{LOS in WINNER+B1 with antenna height = $1.5$ m;}\\
			& \multicolumn{4}{l|}{pathloss at $3$ m is used for distance $<3$ m}\\
			\hline
			\multirow{2}{*}{Shadowing fading} & \multicolumn{4}{l|}{Log-normal distributed with $3$ dB standard deviation,}\\
			& \multicolumn{4}{l|}{and decorrelation distance of $25$ m}\\
			\hline	
		\end{tabular}
	}
\end{table}

We evaluate the performance of VRLS in terms of reliability of the V2V communication in DOCA, and report the policy it develops. We first compare its performance with the state of the art, including the results of our previous work \cite{sahin2018reinforcement}. We then demonstrate its ability to handle the HD constraints, as well as collisions, under different resource pool configurations. Finally, we present the performance VRLS achieves in case of more complex vehicle mobility.

\subsection{Simulation Setup}
We consider a cellular network system with a DOCA assumed to be a straight section of a highway, on which vehicles travel at constant speeds. Details of the simulation parameters related to the vehicular environment, as well as the training of the RL agent are, provided in Table \ref{tableScenario}.

Each simulation starts with a random assignment of resources to the vehicles, and a random action taken by the agent. Moreover, first message generation instance of each vehicle entering DOCA is randomized uniformly across the subframes of the resource pool. 
The policy is trained for each scenario using 16 agents in parallel, each interacting with a different random seed of the simulation environment. The performance of the trained agents is evaluated in the environments having a different random seed from the ones they were trained on. In the results, the performance metric (average PRR) is shown, collected every $10$~s, for a time period containing more than $1000$ resource assignments/actions.

\subsection{Comparison of VRLS and State-of-the-art Algorithms}
\label{Comparison}

\begin{figure}[!t]
	\centering
	\includegraphics[scale=0.53]{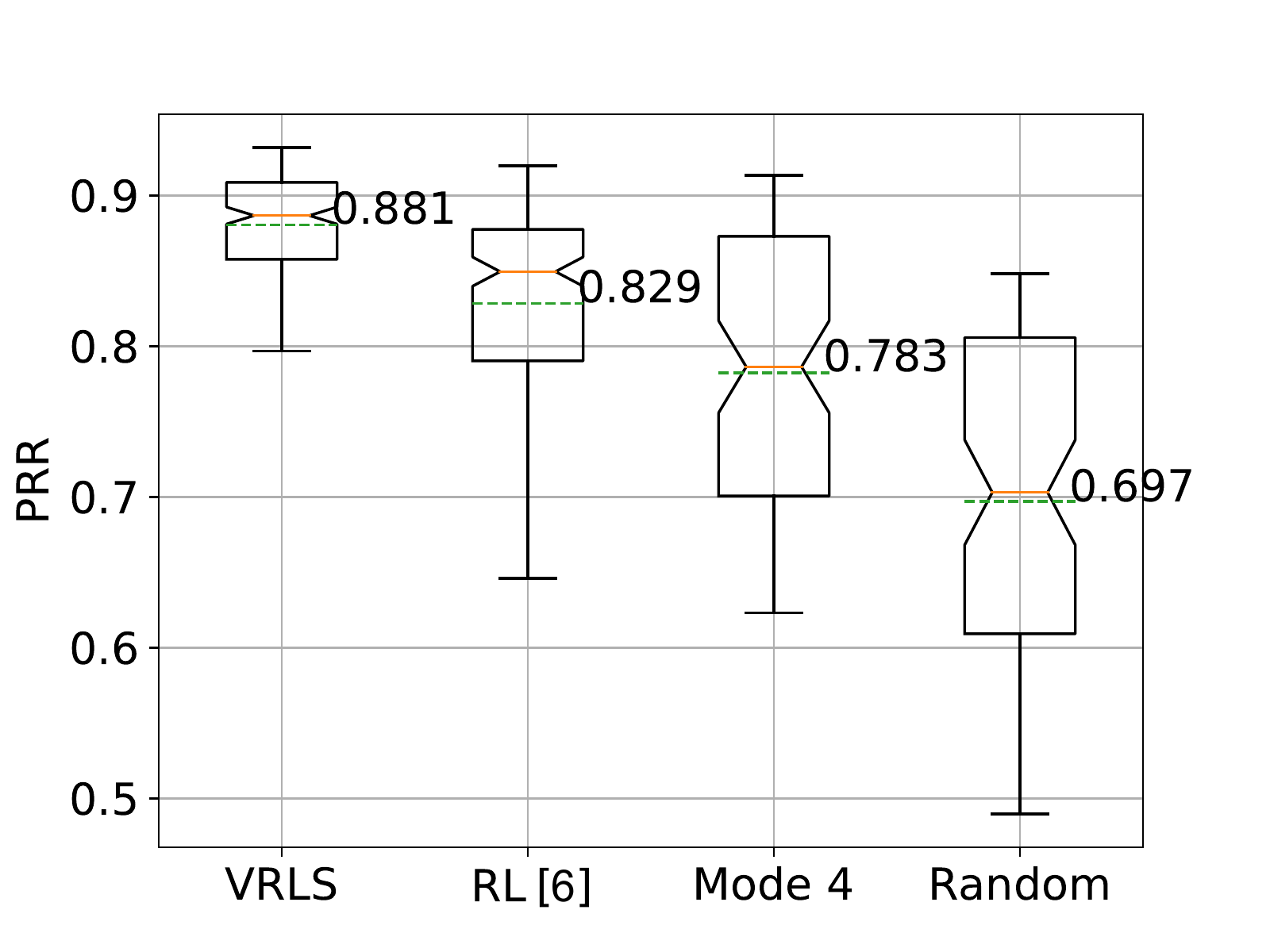}
	\caption{Comparison of VRLS to the state of the art. Mean (green, dashed, denoted), median (red) with $95\%$ confidence interval around (notches), $25^{th}$ and $75^{th}$ percentiles (box), and $1^{st}$ and $99^{th}$ percentiles (whiskers) of PRR.}
	\label{Results_E2}
\end{figure}

We first compare the performance of VRLS with our solution proposed in~\cite{sahin2018reinforcement}, as well as the distributed scheduling algorithm Mode-4 from the 3GPP standard~\cite{3gppTR36885}, autonomously performed by the vehicles based on a sensing mechanism, and the random resource allocation performed by a centralized scheduler, upon each vehicle entering DOCA. The parameter ``probResourceKeep'' of Mode 4 is set to 0, which leads to dynamic reselection of resources as much as possible.
 

We evaluate the performance of the algorithms in an overloaded network scenario with multiple collision domains inside DOCA. To achieve this condition, $30$ vehicles are assumed to be traveling in DOCA (of $500$-m length), with an available resource pool configured with $2$ subchannels by $10$ subframes, where their transmission ranges are limited to $120$~m by adjusting the transmission powers. The vehicular density in DOCA is kept constant by re-inserting vehicles back to DOCA from the opposite direction once they leave it. 

\begin{figure}[!t]
	\centering
	\includegraphics[scale=0.55]{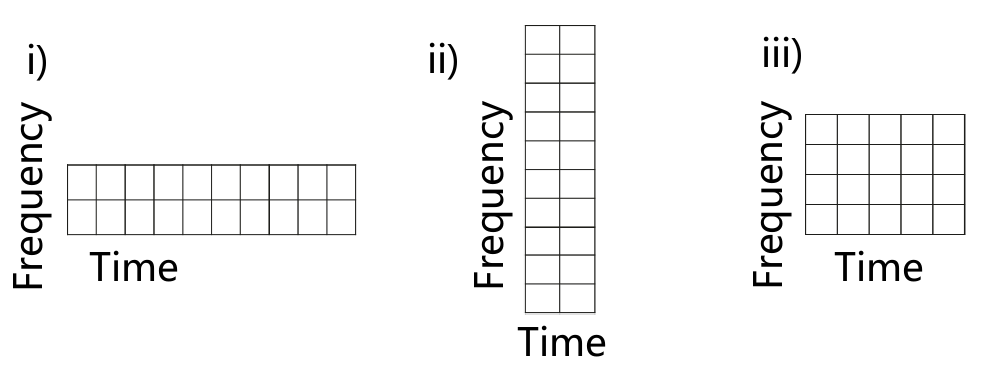}
	\caption{Different configurations of resource pools considered for evaluations in Section \ref{learnHD}.}
	\label{HDpools}
\end{figure}

\begin{figure}[!t]
	\centering
	\includegraphics[scale=0.42]{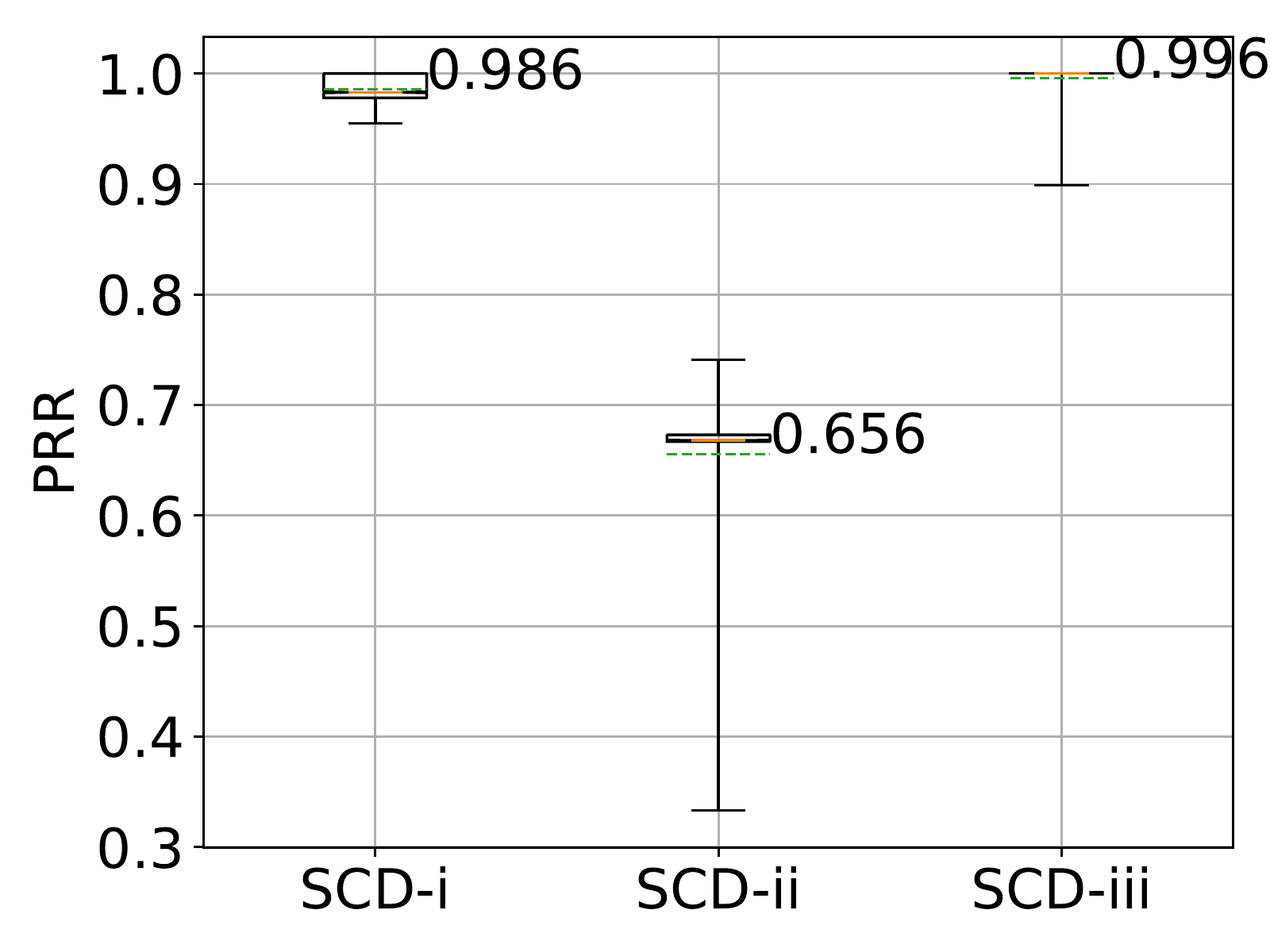}
	\caption{Performance of VRLS on a single-collision-domain (SCD) DOCA, with added complexity to mobility, in scenarios SCD-i, SCD-ii, and SCD-iii. Mean (green, dashed, denoted), median (red) with $95\%$ confidence interval around (notches), $25^{th}$ and $75^{th}$ percentiles (box), and $1^{st}$ and $99^{th}$ percentiles (whiskers) of PRR.}
	\label{Results_HD}
\end{figure}

Furthermore, we introduce the effects of pathloss and shadow fading on the wireless channel to the evaluation environment {\color{black}(with parameters in Table \ref{tableScenario}), which are not present in the pre-training environment of the VRLS and RL agent from \cite{sahin2018reinforcement}. This way, we evaluate the  capability of the RL agents to \textit{transfer learning} to a new environment, modified in terms of wireless channel. We report their performance after a limited re-training (for $200$ epochs) on the new environment.} 
We here note that the upper limit of achievable PRR in this evaluation scenario is analytically intractable to calculate, as it would require the full and instantaneous knowledge of the environment at the scheduler. Accordingly, we compare the reliability performance of VRLS with the state-of-the-art solutions as baselines.

We present the results of the algorithms in Fig. \ref{Results_E2} where PRRs are collected up to $100$~m away from the transmitters. The result indicates the transfer learning capability of VRLS, as it reaches high PRRs, up to $93\%$, and outperforms all of the state-of-the-art solutions. Specifically, we observe a significant improvement on low percentiles. 
This performance gain is mainly achieved by the difference in state representation, as described in Section \ref{State}. VRLS achieves such performance after only $800$ epochs of training over our training environment, and with some limited retraining over the evaluation environment, up to $200$ epochs. 
 
In terms of the developed policy, VRLS agent learns to divide the resource pool dynamically into two directions of the highway, proportional to the density of each direction, while performing resource reuse per direction. This way, resources are efficiently utilized while aiming to minimize the collisions, with the trade-off controlled by the received award. On the other hand, HD errors occur due to agent's allocations, which in this scenario is unavoidable given the overloaded conditions of the network. In majority of such cases, subchannels sharing the same subframe are assigned to vehicles moving in \textit{opposite} directions. Such vehicles would not be able to listen to each other when passing each other for a short duration of time. However, this type of allocation degrades the PRR to a lesser extent compared to the impact of alternative policies, e.g., HD errors or collisions that would otherwise occur more persistently in the same direction.

\subsection{Learning the Half-duplex Constraint}
\label{learnHD}

In the previous subsection, we study how VRLS performs in an overloaded MCD scenario, demonstrating its capability to reuse TBs, and prevent collisions. In this subsection, we evaluate the performance of VRLS in scenarios that specifically require its capability of learning and solving the HD constraint, 
given different configurations of the resource pool, as shown in Fig. \ref{HDpools}. We consider underloaded network conditions in a single-collision-domain (SCD) DOCA, where all vehicles inside are able to sense each other's transmissions, and any resource reuse leads to collision. Fig. \ref{HDpools} shows the three resource pool configurations. The first two scenarios represent the two extremes of a resource pool configuration: SCD-i) $10$ subframes by $2$ subchannels, and SCD-ii) $2$ subframes by $10$ subchannels. For SCD-i and SCD-II, we simulate a maximum number of $10$ and $4$ vehicles, respectively. The third scenario (SCD-iii) lies in between:  $5$ subframes by $4$ subchannels, and we simulate $5$ vehicles. These scenarios are chosen such that any HD errors would decrease the PRR considerably, and the optimal resource allocation is possible only if the HD relation among the resources is learned by the scheduler. 
Differing from the previous evaluations, we also introduce an added complexity to the vehicular mobility, where after leaving, vehicles are re-inserted to DOCA after a time offset distributed exponentially at random with a $2.5$~s mean, which introduces time-varying vehicular density inside DOCA.

The performance of the RL agent for each scenario is shown in Fig.~\ref{Results_HD} in terms of PRR measured over the entire DOCA (i.e., within $500$~m distance). We observe that VRLS can easily adapt to each of the settings, 
and performs near optimal in all scenarios (i.e., close to the analytical maximum), {\color{black}after a training of around $500$ epochs}. 
In SCD-i and SCD-iii, $100\%$ PRR is achievable analytically, if the TBs assigned to vehicles are all orthogonal in time (i.e., chosen from different subframes). In SCD-ii, in case all four vehicles are inside the DOCA, then in the best case, two vehicles are assigned different subchannels in one subframe and two in another, yielding a PRR of $66.7\%$ (limited by the HD constraint). Higher PRRs are achievable in the case of fewer number of vehicles traveling through DOCA. 


In SCD-i, a {\color{black} single} HD error due to {\color{black} an} assignment of two resources non-orthogonal in time would {\color{black} analytically} lead to $97.7\%$ PRR, which {\color{black} can be observed} in around $25\%$ of the cases. In SCD-ii, a single HD conflict would result in a PRR of $50\%$, observed in less than $25\%$ of the cases. In case of SCD-iii, the agent is able to achieve a similar performance, having a single HD conflict in $1\%$ of the cases, which results in a PRR of $90\%$. Moreover, {occasionally \color{black}in SCD-ii, there are fewer than four vehicles traveling in DOCA}, 
where a {\color{black}single} HD error between two vehicles would yield a PRR of $0\%$. The trained RL agent is {\color{black}successfully} able to yield non-zero PRRs more than $99\%$ of the time.  

On the other hand, compared to HD errors, any collision error (due to assignment of the same TB to more than a single vehicle) would reduce the PRR to a greater extent. As an example, in SCD-i, assigning the same TB to a single pair of vehicles in DOCA would result in an analytically derived PRR of $80\%$. Such cases were only observed in less than $1\%$ of the time, which shows the success of the RL agent on avoiding the collisions.
Overall, the results show the ability of VRLS to learn and deal with the HD constraint, in addition to avoiding the collisions, achieved in three different resource pool configurations. 

\subsection{Multi-collision-domain DOCA}

\begin{figure}[!t]
	\centering
	\includegraphics[scale=0.4]{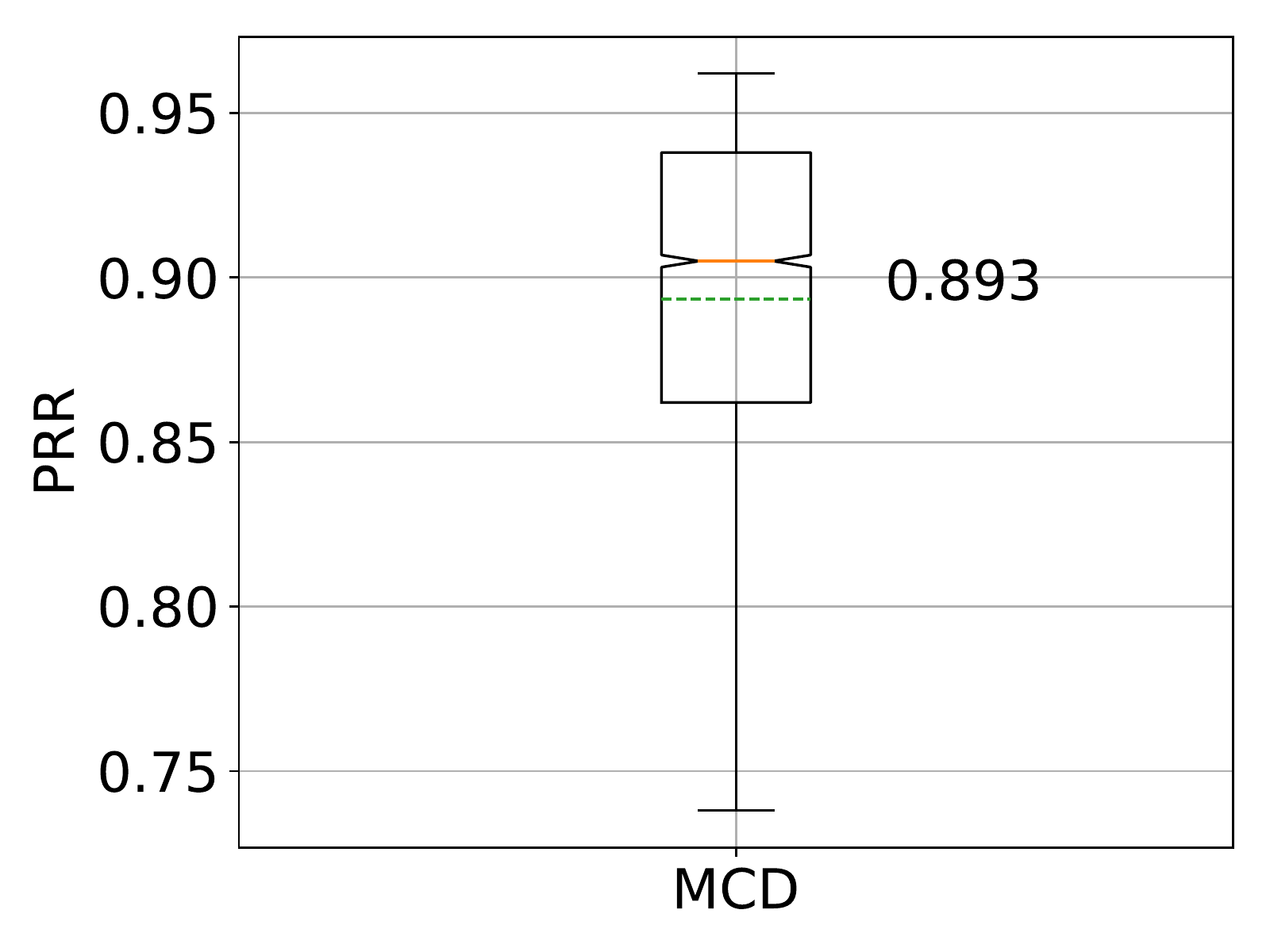}
	\caption{Performance of VRLS on a multi-collision-domain (MCD) DOCA, with added complexity to mobility. Mean (green, dashed, and denoted), median (red) with $95\%$ confidence interval around (notches), $25^{th}$ and $75^{th}$ percentiles (box), and $1^{st}$ and $99^{th}$ percentiles (whiskers) of PRR.}
	\label{Results_E2plus}
\end{figure}

Lastly, we evaluate the performance of VRLS in a multi-collision domain scenario {\color{black} with added complexity in terms of mobility}. Specifically, we consider the same scenario as in Section \ref{Comparison} {\color{black} without the pathloss and fading effects}, however with a time-varying vehicular density in DOCA. The results are provided in Fig. \ref{Results_E2plus}, which shows that the RL agent is able to achieve a performance comparable with the VRLS performance in Fig. \ref{Results_E2}, which has a simpler mobility. On the other hand, the performance shows higher variance that could be seen by looking at the difference between the $1^{st}$ and $25^{th}$ percentiles of PRR. The policy observed is similar to the case in Section \ref{Comparison}, which dynamically divides the resources into directions, and could be regarded as desirable considering the highly varying density of vehicles per direction over time.


\subsection{On Real-world Deployment of VRLS}
In this study, we consider a simulation-based approach for training and evaluating the VRLS. In a real-world deployment of VRLS, the state information could be easily monitored at the BSs, by keeping the record of how long ago a resource is allocated, how many times, and to which direction. VRLS could be pre-trained on a realistic simulation environment, and further fine-tuned after the deployment by exploiting any feedback provided by the vehicles coming \textit{back} to coverage.

	\section{Conclusions and Future Work}
\label{Conclusion}

We proposed VRLS, a unified scheduling approach for V2V communications based on RL, and showed that it outperforms the state-of-the-art V2V scheduling algorithms by:
i) learning about the collisions in case of non-orthogonal resource assignment to nearby vehicles; and
ii) learning that half duplex (HD) constraint needs to be accounted for. VLRS is designed by unifying the state, reward, and action definitions so that it can adapt well to variations in the vehicular environment ({\color{black}  in terms of} vehicle density), radio propagation conditions, and different resource pool configurations. 

Vehicular communications environment has a wide set of variables, which makes scheduling V2V communication a challenge. In this paper we focused on 
some salient aspects such as HD constraint, resource pool configuration in time and frequency, vehicle density, etc. In the follow-up work, we will evaluate VRLS by considering vehicles traveling at varying speeds, different road configurations, and the possibility to combine pre-scheduling with dynamic resource selection within the DOCA.

	\bibliographystyle{IEEEtran}

\end{document}